\documentclass[11pt]{iopart}

\usepackage{iopams}
\usepackage{epsfig}
\usepackage{amssymb}
\usepackage{graphicx}
\usepackage{geometry}
\usepackage{color}

\date{\today}

\usepackage{textcomp}
  \expandafter\let\csname equation*\endcsname\relax
  \expandafter\let\csname endequation*\endcsname\relax
\usepackage{amsmath}
\usepackage{amsfonts}
\usepackage{amsthm}

\theoremstyle{plain}
\newtheorem{thm}{Theorem}[section]

\newtheorem{lem}[thm]{Lemma}
\newtheorem{prop}[thm]{Proposition}

\theoremstyle{definition}
\newtheorem{defn}{Definition}[section]

\theoremstyle{remark}
\newtheorem{ex}{Example}
\theoremstyle{remark}
\newtheorem{oss}{Remark}

\newcommand{\Z}{\mathbb{Z}}
\newcommand{\hX}{\widehat{X}}

\begin{document}

\title[Slicer map and Lev\'y walks]{A trivial non-chaotic map lattice asymptotically 
indistiguishable from a L\'evy walk}
\author{Lucia Salari$^1$, Lamberto Rondoni$^1$, Claudio Giberti$^2$}
\address{$^1$Dipartimento di Scienze Matematiche, Politecnico di Torino,
Corso Duca degli Abruzzi 24 I-10129 Torino, Italy\\ 
$^2$Dipartimento di Scienze e Metodi dell'Ingegneria, Universit\'a di Modena e Reggio E.,
Via G. Amendola 2 - Pad. Morselli, I-42122 Reggio E., Italy  }
\eads{ \mailto{lucia.salari@polito.it}, \mailto{lamberto.rondoni@polito.it},\\ \mailto{claudio.giberti@unimore.it}}
\begin{abstract}
In search for mathematically tractable models of anomalous diffusion, we introduce a simple dynamical system consisting
of a chain of coupled maps of the interval whose Lyapunov exponents vanish everywhere. The volume preserving property 
and the vanishing Lyapunov
exponents are intended to mimic the dynamics of polygonal billiards, which are known to give rise to anomalous
diffusion, but which are too complicated to be analyzed as thoroughly as desired. Depending on the value 
taken by a single parameter $\alpha$, our map experiences sub-diffusion, super-diffusion or normal diffusion. 
Therefore its transport properties can be compared with those of given L\'evy walks describing transport in 
quenched disordered media. Fixing $\alpha$ so that the mean square displacement generated by our map and that
generated by the corresponding L\'evy walk asymptotically coincide, we prove that all moments of the 
corresponding asymptotic distributions coincide as well, hence all observables which are expressed in terms 
of the moments coincide.\\ \\
{\bf Short title:} { Non-chaotic maps and  L\'evy walks }
\end{abstract}
\submitto{\NL}
\maketitle
\tableofcontents

\section{Introduction}
The problem of anomalous or nonlinear diffusion is fundamental in many applications, e.g.\ its recent 
applications to osmotic-like phenomena \cite{IGARASHI2011}. When the size of solute molecules 
becomes smaller but still comparable with pore size, their flow tipically becomes anomalous 
\cite{JR2006,JRB2008,FDT2008,JR2010,JEPPS2003,BATHIA2011} and may lead to 
a sequence of quasi-stationary states in which the amount of solute on both sides of the membrane remains practically 
constant for very long periods of time. As such applications are presently object of intense theoretical investigation 
and technological development, it is important to find minimalistic models in which the essential ingredients of
such complicated processes can be examined in detail.

Among the models of transport in pores of size comparable to that of the transported molecules, we find 
the polygonal billiards \cite{Gutkin,JR2006,JRB2008}, whose dynamics are not chaotic, in the sense 
that their Lyapunov exponents vanish. These billiards enjoy a wide range of transport 
properties, but are very hard to understand as in detail as desired. Therefore, analogously to
idealizations of convex billiards consisting of chaotic maps yielding standard diffusion
\cite{GASPARD,DORFMANN,VOLLMER}, we introduce simple volume preserving non-chaotic maps, in 
search for the minimal ingredients leading to the anomalous transport common to polygonal billiards.

Let us recall that standard transport generated by chemical potential gradients is described by 
Fick's first law \cite{GM1984}:
\begin{equation}\label{eq:FICK1}
J(x)=-D\frac{\partial c}{\partial x}
\end{equation}
where $J$ is the mass flow, $D$ the diffusion coefficient, $c$ the concentration and $x$ the position 
in space. This law, which can be justified in kinetic theory \cite{CC1970}, leads to Fick's second law:
\begin{equation}\label{eq:FICK2}
\frac{\partial c}{\partial t}= D \frac{\partial^{2}c}{\partial x^{2}}
\end{equation}
where $t$ is the time variable, whose solution for an initial $\delta$-function distribution is given by:
\begin{equation}\label{eq:FICKSOL}
c(x,t)=(4\pi Dt)^{-1/2} e^{-x^{2}/4Dt} 
\end{equation}
so that
\begin{equation}\label{eq:FICKCOEFF}
\langle x^{2}(t)\rangle = \int_{-\infty}^{\infty} x^{2} c(x,t)dx = 2Dt
\end{equation}
Therefore, it is customary to call {\em diffusive} any phenomenon enjoying
a linear relation such as (\ref{eq:FICKCOEFF}), even if only asymptotically in time. 
In general, one refers to the following:
\begin{defn}
\textit{Let $\langle \Delta x^{2}(t)\rangle$ be the mean square displacement, and introduce
\begin{eqnarray} \label{eq:GENDIFFCO}
D_\gamma :=\lim_{t \to \infty} \frac{\langle \Delta x^{2}(t)\rangle}{t^{\gamma}}~.
\end{eqnarray}
If $D_\gamma\in(0,\infty)$ for a value $\gamma \in [0,2]$, $\gamma$ is called transport exponent,
$D_\gamma$ is called generalized diffusion coefficient, and the transport is called:
\begin{enumerate}	
	\item[(i)] sub-diffusive if $\gamma \in [0,1)$
	\item[(ii)] diffusive if $\gamma=1$ 
	\item[(iii)] super-diffusive if $\gamma \in (1,2]$. It is also called ballistic for 
$\gamma=2$~\footnote{$\gamma$ cannot be larger than 2, since the travelled distance
cannot exceed the largest speed multiptied by $t$.}
\end{enumerate}}
\end{defn}
In the following we introduce a trivial coupled map lattice which enjoys all the above kinds of 
diffusion in one dimension, depending on a parameter that must be fixed. This dynamical system 
turns out to be indistinguishable from certain L\'evy walks, as far as the
asymptotic behaviour of its moments is concerned. We call {\em Slicer} this dynamical system.
 
The paper is organized as follows. In Section 2, we define the Slicer model and describe its transport 
properties, some example of which are reported in Section 3. In Section 4, we compare 
the Slicer dynamical system with the L\'evy walks in a quenched disordered media  studied in  
\cite{BURIONI2010}. Section 5 is devoted to concluding remarks.

\section{The Slicer Dynamics}
Consider the unit interval $M:=[0,1]$, the chain of such intervals $\widehat{M}:=M\times \mathbb{Z}$,
and the product measure $\hat{\mu}:=\lambda\times\delta_{\mathbb{Z}}$ on $\widehat{M}$, where $\lambda$  
is the Lebesgue measure on $M$ and $\delta_{\mathbb{Z}}$ is the Dirac measure on the integers.
Denote $\pi_{M}$ and $\pi_{\mathbb{Z}}$ the projections of $\widehat{M}$ on its first and second factors.
Let $x$ be a point in $M$, $\hX=(x,m)$ a point in $\widehat{M}$, and  
$\widehat{M}_{m}:=M\times\left\{m\right\}$ the $m$-th cell of $\widehat{M}$. Subdivide each $\widehat{M}_{m}$
in four sub-intervals, separated by three points {\em ``slicers''}:
$$\left\{ 1/2 \right\}\times\left\{m\right\}~, \quad
\{\ell_m\}\times\left\{m\right\}~, \quad 
\{1-\ell_m\}\times\left\{m\right\}~, 
$$ 
where $0 < \ell_m < 1/2$, for every $m \in \mathbb{Z}$. 

{\em Slicer} is a dynamical system $(\widehat{M},\hat{\mu},S)$ which, at each time step $n\in\mathbb{N}$, 
moves all sub-intervals from their cells to neighbouring cells, implementing the rule  
$S : \widehat{M} \to \widehat{M}$ defined by:  
{ 
\begin{equation}\label{eq:MAPPAGEN}
S(x,m)= 
\left\{
\begin{array}{rl}
(x,m-1) & \mbox{ if } 0\leq x <  \ell_{m} \mbox{ or } \frac{1}{2} < x \leq 1-\ell_{m},\\
(x,m+1) & \mbox{ if } \ell_{m} < x \leq \frac{1}{2} \mbox{ or } 1-\ell_{m} \leq x \leq 1,
\end{array}
\right.
\end{equation}
This map is invertible, with inverse given by
\begin{equation}\label{eq:INVMAPPAGEN}
S^{-1}(x,m)= 
\left\{
\begin{array}{rl}
(x,m+1) & \mbox{ if } 0\leq x <  \ell_{m} \mbox{ or } \frac{1}{2} < x \leq 1-\ell_{m},\\
(x,m-1) & \mbox{ if } \ell_{m} < x \leq \frac{1}{2} \mbox{ or } 1-\ell_{m} \leq x \leq 1 ~.
\end{array}
\right.
\end{equation}  
It is also time reversal invariant (TRI), i.e.\ there exists an involution $i$
such that $ i S=S^{-1}i$, e.g.\ the one defined by $i(x,m)=(1-x,m)$ will do. The action of $S$ is illustrated by 
Fig.\ref{fig:SLICER}, for an initial condition concentrated and uniform in $\widehat{M}_{m}$.
\begin{figure}[h]
\centering
\includegraphics[scale=0.7]{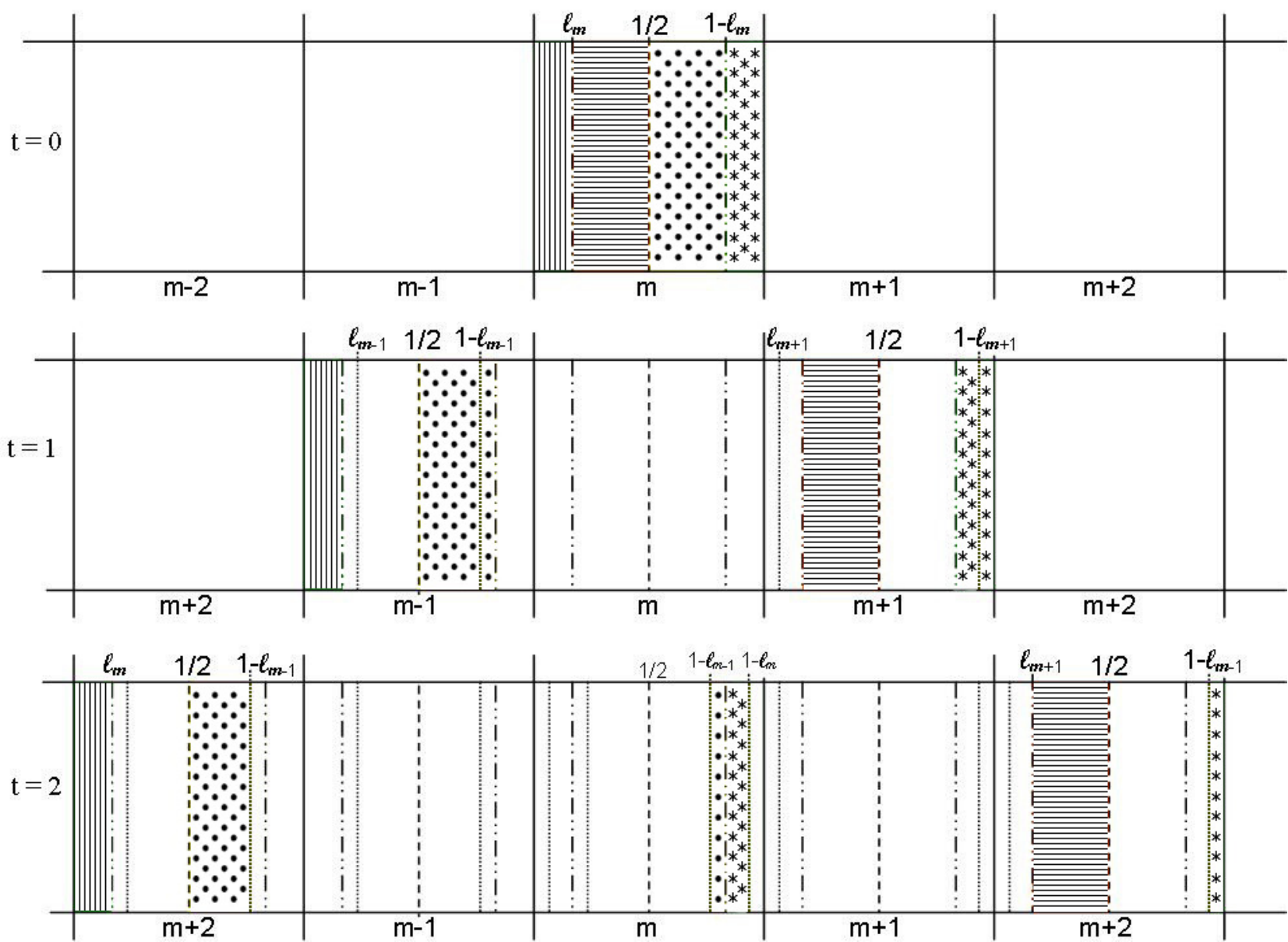}
\caption{Two time steps of the slicer map, for initial conditions in cell $\widehat{M}_{m}$.  
For the sake of the figure a vertical coordinate has been added.}
\label{fig:SLICER}
\end{figure}

For every $\alpha > 0$, let us introduce the family of slicers:
\begin{equation} \label{eq:SLICEGEN}
L_{\alpha}=\left \{\Big(\ell_{j}(\alpha), 1-\ell_j(\alpha)\Big) : 
\ell_j(\alpha) = \frac{1}{\left(\left|j\right|+2^{1/\alpha}\right)^{\alpha}},\, j\in \mathbb{Z}\right\}
\end{equation}
Then, the slicer map is denoted by $S_\alpha$ if all slicers of Eq.(\ref{eq:MAPPAGEN}) belong to $L_\alpha$: 
$\ell_m=\ell_m(\alpha)$. Obviously, for every $\alpha>0$, $S_\alpha$ preserves $\hat{\mu}$ and is not chaotic 
(its Lyaponuv exponents vanish). Indeed, different points in $\widehat M$ neither 
converge nor diverge from each other in time, except when separated by a slicer, in which case their distance 
jumps. But like for two particles in a polygonal billiard, one of which hits a corner of the polygon while the 
other continues its free flight \cite{JR2006}, the separation points constitute a set of zero 
$\hat{\mu}$ measure, hence they do not suffice to produce positive Lyapunov exponents.


The transport  properties of the Slicer Dynamics will be examined by taking an ensemble of points  
$\widehat E_0$ in the central cell $\widehat{M}_0=M\times \left\{0\right\}$ and studying 
the way $S_{\alpha}$ spreads them in $\widehat{M}$. One finds 
that in $n$ time steps the points of $\widehat E_0$ reach $\widehat{M}_n$ and $\widehat{M}_{-n}$,
and that the cells occupied at time $n$ have
odd index if $n$ is odd, and have even index if $n$ is even, as illustrated in 
Fig.\ \ref{fig:MAPPA7FIG} for a given $\alpha$.
\begin{figure}[h]
\centering
\includegraphics[scale=0.7]{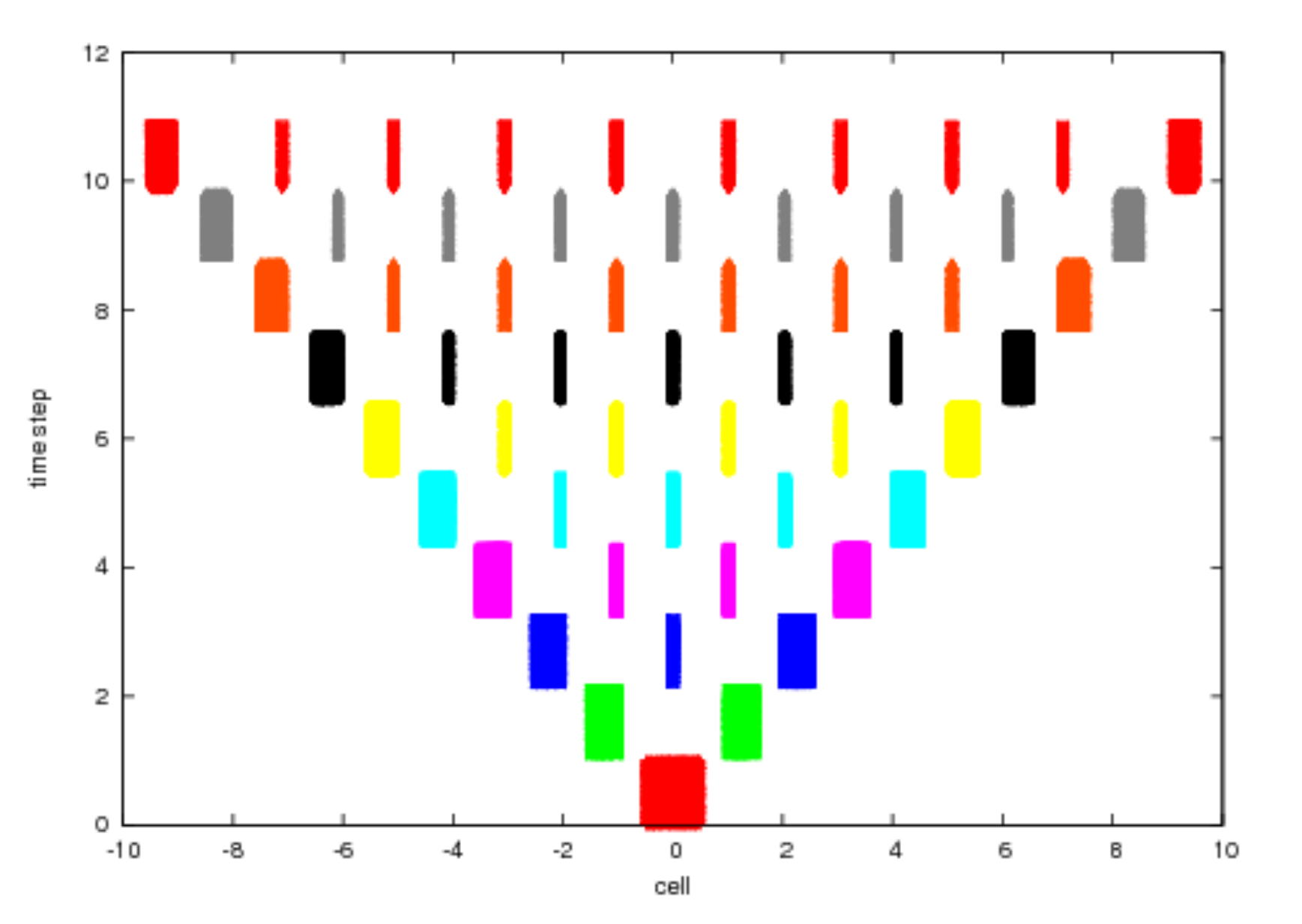}
\caption{Ten time steps of $S_\alpha$, with $\alpha=1/3$, for $N=10^{4}$ points initially uniformly
distributed in $\widehat{M}_0$. The direction of time is the vertical axis. For the sake of the figure, an extra inactive vertical 
coordinate has been added.}
\label{fig:MAPPA7FIG}
\end{figure}
More precisely, taking 
\begin{equation}\label{eq:PnDn}
 P_{n}=\{ j\in \Z :  j\  \mbox{is even and\ }  |j|\le n\},\quad 
D_{n}=\{ j\in \Z :  j\  \mbox{is odd and\ }  |j|\le n\},
\end{equation}
we have
\begin{equation}
S_{\alpha}^{n}~\widehat{M}_{0}=\bigcup_{j\in P_{n}} \big(R_{j}\times \{j\}\big) \quad \mbox{if~} n \mbox{~is even}~,
\qquad 
S_{\alpha}^{n}~\widehat{M}_{0}=\bigcup_{j\in D_{n}} \big(R_{j}\times \{j\}\big),\quad \mbox{if~} n \mbox{~ is odd}
\end{equation}
where $R_{j}\times \{j\}\subset \widehat{M}_{j}$, and $R_{j} \subset M$ is an interval or a union of intervals 
if $\widehat{E}_{0}=\widehat{M}_{0}$.

Let $d \nu_{0}:=\hat{\rho}_{0}(\widehat{X}) d \hat{\mu}$ be a probability measure on $\widehat M$ with density
\begin{equation}\label{eq:PDF0}
\hat{\rho}_{0}\left(\widehat{X}\right)=
\left\{
\begin{array}{rl}
1, \ \ &\mbox{if } \widehat{X}\in M_{0}\\
0, \ \ &\mbox{otherwise}
\end{array}
\right.
\end{equation} 
This measure evolves under the action of $S_\alpha$ describing the transport of an ensemble of points 
initially uniformly distributed in $\widehat{M}_{0}$. For simplicity, in the following we will always 
adopt this initial condition, which mimics the $\delta$-function initial condition common in standard 
diffusion theory; if the initial condition were confined within $\widehat{M}_m$ with $m \ne 0$, nothing 
would change in practice. 

Requiring the conservation of probability, the evolution $\nu_n$ of $\nu_{0}$ 
at time $n$ is given by $\nu_{n}(\widehat{R})= \nu_{0}(S^{-n}_{\alpha}\widehat{R})$, for every measurable $\widehat{R} \subset \widehat{M}$.
Its density is given by:
\begin{equation}\label{eq:PDFN}
\hat{\rho}_{n}(\widehat{X})=
\left\{
\begin{array}{rl}
1,\ \ \  & \mbox{if } \widehat{X}\in S_{\alpha}^{n} \widehat{M}_{0} ~~~ \left(\mbox{or, equivalently, if ~} 
S_{\alpha}^{-n} \widehat{X}\in \widehat{M}_{0} \right) \\
0, \ \ \  & \mbox{otherwise}.
\end{array}
\right.
\end{equation}
Consider the sets
\begin{equation}
\widehat{R}_{j}:=S^{n}_{\alpha} (\widehat{M_{0}}) \cap {\widehat{M}_{j}},\quad j=-n,\ldots, n,
\end{equation}
which constitute the total phase space volume occupied at time $n$ in cell ${\widehat{M}_{j}}$. Their
measure
\begin{equation}
A_{j}:= \hat{\mu}( \widehat{R_{j}} )=\lambda(\pi_{M}(\widehat{R_{j}}))\delta_{\mathbb{Z}}(j)=\lambda(\pi_{M}(\widehat{R_{j}})),
\end{equation}
equals the probability $\nu_n(\widehat{M}_{j})$ of cell $j$ at time $n$: as $\hat{\mu}$ is invariant and
$S_{\alpha}$ invertible, we have
$$
A_{j}=\hat{\mu}( \widehat{R}_{j})=\hat{\mu}( S^{-n}_{\alpha}(\widehat{R}_{j}))
=\hat{\mu}(\widehat{M_{0}}\cap S^{-n}_{\alpha}(\widehat {M_{j}}))=\nu_{0}(S^{-n}_{\alpha}(\widehat {M_{j}}))
=\nu_{n}({\widehat{M}_{j}}) ~,
$$ 
and 
$\sum_{j=-n}^{n} A_{j} = \hat{\mu}(\cup_{j=-n}^{n} S^{n}_{\alpha} (\widehat{M}_{0}) \cap {\widehat{M}_{j}} )=
\widehat{\mu}(S^{n}_{\alpha} (\widehat{M_{0}}) )=\hat{\mu}(\widehat{M_{0}}) = 1$. Indeed,
$S^{n}_{\alpha} (\widehat{M_{0}}) \cap {\widehat{M}_{j}} =\emptyset$ for $|j|>n$ and $\cup_{j=-\infty}^{\infty } {\widehat{M}_{j}} = \widehat{M}$.
In other words, the $A_{j}$'s define a probability distribution which coincides with $\nu_{n}(\pi_{\mathbb Z}^{-1})$ 
and, thus, is a marginal probability distribution of $\nu_{n}$. 
Starting from the ``microscopic'' distribution $\nu_n$ on $\widehat{M}$, we can now introduce its coarse grained version
$\rho_{n}^{G}$, as the following measure on the integer numbers $\mathbb{Z}$:
\begin{defn}
\textit{For every time $n\in\mathbb{N}$, the Coarse Grained Distribution is defined by
\begin{equation}
\rho^{G}_{n}(j)=
\left\{
\begin{array}{rl}
A_j \ \ & \mbox{if } j\in \{-n,\ldots, n \},\ \\
0 \ \ \  & \mbox{\rm otherwise.}
\end{array}
\right.
\end{equation}
$A_{-n}$ and $A_{n}$ are called traveling areas; $A_j$ is called sub-traveling area
if $|j|<n$.
}
\end{defn}
\begin{oss}
From the definition of $S_{\alpha}$ and the initial condition (\ref{eq:PDF0}),
we have $A_{j}=A_{-j}$ for all $j\in \mathbb{Z}$. Thus, $\rho^{G}_{n}(j)$ is
even, $\rho^{G}_{n}(j)=\rho^{G}_{n}(-j)$, and all its odd moments vanish. 
\end{oss}

The Coarse Grained Distribution will be used to describe the transport properties of the coarse grained trajectories  
$\{\pi_{\mathbb{Z}}(S^{n}_{\alpha}(\widehat{X}_{0}))\}_{n\in \mathbb{N}}\subset \mathbb{Z}$, with 
$\widehat{X}_{0} \in \widehat{M}_{0} $. 
This way, $\rho_{n}^{G}$ becomes the discrete analogous of the mass concentration $c(x,t)$ of 
Eqs.(\ref{eq:FICK1}) and (\ref{eq:FICK2}), and
we can introduce the discrete versions of Eqs.(\ref{eq:FICKCOEFF}) and (\ref{eq:GENDIFFCO}) as follows:
\begin{defn}
\textit{The mean square displacement induced at time $n$ by $S_\alpha$ is the second moment
\begin{equation}\label{eq:DMSD}
\langle \Delta \hat{X}^{2}_{n}\rangle := \sum_{j=-n}^{n} A_j j^{2}
\end{equation}
of $\rho_{n}^{G}$, where $j$ is the distance travelled by a point in $\widehat{M}_j$ at time $n$.  
Then, for $\gamma \in [0,2]$ let
\begin{equation}
T_{\alpha}(\gamma):=\lim_{n\rightarrow \infty} \frac{\langle \Delta \hat{X}^{2}_{n}\rangle }{n^{\gamma}} ~.
\end{equation}
If $T_{\alpha}(\gamma^t)\in (0,\infty)$ for $\gamma^{t}\in [0,2]$, $\gamma^{t}$ is called the  
transport exponent of the Slicer Dynamics, and  $T_{\alpha}(\gamma^{t})$ the generalized
diffusion coefficient.
} 
\end{defn}  
\begin{oss}
The mean displacement  $\langle \Delta \hat{X}_{n}\rangle := \sum_{j=-n}^{n} A_j j$  vanishes at all $n$, 
hence there is no drift in the Slicer Dynamics. \\
\end{oss}
Note that $A_{j}$ equals the width of the interval $R_{j}$, which is determined by the position of the slicers in the 
$j$-th cell, once $\alpha$ is given. Therefore, $A_{j}$ can be computed directly from Eq.(\ref{eq:SLICEGEN}). 
For the traveling areas we have
\begin{equation}\label{eq:AV}
A_n=\ell_{n-1}=\left(\frac{1}{|n|-1+2^{1/\alpha}}\right)^{\alpha} = A_{-n}
\end{equation}
while for the non vanishing sub-traveling areas we have:
\begin{equation}
A_j = \ell_{\left|j\right|-1}-\ell_{\left|j\right|+1} =
\frac{1}{\left(\left|j\right|-1+2^{1/\alpha}\right)^{\alpha}}-
\frac{1}{\left(\left|j\right|+1+2^{1/\alpha}\right)^{\alpha}}~,
\end{equation}
For even $n>2$, this implies
\begin{equation}\label{eq:DENSFUNKP}
{\rho}^{G}_{n}(j)=
\left\{
\begin{array}{rl}
2(\ell_{0}-\ell_{1}) ~,       & \ \ \mbox{for }  j=0\\
\ell_{\left|2k-1\right|}-\ell_{\left|2k+1\right|} ~, & \ \ \mbox{for } |j|=2k,\, \ k=1,\ldots,\frac{n-2}{2}\\
\ell_{\left|n-1\right|} ~,           & \ \ \mbox{for } |j|=n\\
0 ~,                & \ \ \mbox{elsewhere}
\end{array}
\right.
\end{equation}
while for odd $n>3$, it implies:
\begin{equation}\label{eq:DENSFUNKD}
{\rho}^{G}_{n}(j)=
\left\{
\begin{array}{rl}
\ell_{\left|2k\right|}-\ell_{\left|2k+2\right|}~, & \ \ \mbox{for }|j|=2k+1,\ k=0,\ldots,\frac{n-3}{2}\\
\ell_{\left|n-1\right|} ~,         & \ \ \mbox{for } |j|= n\\\
0 ~,                & \ \ \mbox{elsewhere}
\end{array}
\right.
\end{equation}

\begin{oss}
Using Eq.(\ref{eq:SLICEGEN}) in Eqs.(\ref{eq:DENSFUNKD}) and (\ref{eq:DENSFUNKP}) for large $n$, one obtains that
the tail of the distribution (large $j$) goes (independently of the parity of $n$) like: 
${\rho}^{G}_{n}(j)\sim {\displaystyle 2\alpha}/{\displaystyle
|j|^{\alpha+1}}{\mathbb I}_{\{|j|<n\}}$, i.e.\ ${\rho}^{G}_{n}$ has heavy tails.  
In particular, for $\alpha=1/2$, ${\rho}^{G}_{n}$ has tails falling like those of the L\'evy 
distribution, $|j|^{-3/2}$. For $j=\pm n$ the probability is much larger: 
${\rho}^{G}_{n}(n)\sim {\displaystyle  1}/{ \displaystyle  |n|^{\alpha}}$.
\end{oss}

Now we may  state the following:
%
%
\begin{prop}\label{thm:msd}
Given $\alpha\in\left(0,2\right)$ and the uniform initial distribution in $\widehat{M}_0$, 
we have:

\begin{equation} \label{eq:GDIFFCO1}
T_{\alpha}(\gamma)=
\left\{
\begin{array}{rl}
+\infty               & \mbox{if }  0\leq\gamma<2-\alpha \\
\frac{4}{2-\alpha}    & \mbox{if }  \gamma = 2-\alpha    \\
0                     & \mbox{if }  2-\alpha<\gamma\leq 2               
\end{array}
\right. 
\end{equation}
hence the transport exponent $\gamma^{t}$ takes the value $2-\alpha$.
\end{prop}

\proof
For the symmetry of $\rho^{G}_{n}$, it suffices to consider the cells $\widehat{M}_j$ with $j \in \mathbb{N}$:
\begin{equation}\label{eq:AV+AQP}
T_{\alpha}(\gamma)=2 \lim_{n \to \infty} \frac{1}{n^{\gamma}}\sum_{j=0}^{n} A_j j^{2}
= 2 \lim_{n \to \infty} \frac{1}{n^{\gamma}}\left(\sum_{j=0}^{n-1} A_j j^{2} + A_n n^{2}\right).
\end{equation}
Let us focus on the sub-travelling areas first. Lemma \ref{lem:LEMMA1} proved below states that:
\begin{equation} \label{eq:LIMAQP}
\lim_{n \to \infty} \frac{1}{n^{\gamma}}\sum_{j=0}^{n-1} A_j j^{2}=
\left\{
\begin{array}{rl}
\infty \              & \ \ \mbox{if } 0\leq\gamma<2-\alpha \\
\dfrac{\alpha}{2-\alpha}  & \ \ \mbox{if } \gamma = 2 - \alpha \\
0                     & \ \ \mbox{if } 2 - \alpha < \gamma \leq 2
\end{array}
\right.
\end{equation}
Because of Eq.(\ref{eq:AV}), the travelling area yields:
\begin{equation}
\lim_{n\to \infty}\frac{n^{2}}{(n+2^{1/\alpha}-1)^{\alpha}}\cdot \frac{1}{n^{\gamma}}=
\left\{
\begin{array}{rl}
 \infty      &       \mbox{if } 0\leq\gamma<2-\alpha \\
1             &       \mbox{if } \gamma = 2-\alpha \\
0             &       \mbox{if } 2-\alpha<\gamma\leq2
\end{array}
\right.
\end{equation}
Thus, $T_{\alpha}(\gamma)$ is given by:
\begin{equation}
T_{\alpha}(\gamma) = 
\left\{
\begin{array}{rl}
\infty                 &      \mbox{if } 0\leq\gamma<2-\alpha \\
\frac{4}{2-\alpha}     &      \mbox{if } \gamma = 2-\alpha    \\
0                      &      \mbox{if } 2-\alpha<\gamma \leq 2
\end{array}
\right.
\end{equation}
As $\gamma^{t}$ is the value $\gamma$ for which $T_{\alpha}(\gamma)$ is finite, we have $\gamma^{t}=2-\alpha$. 
\endproof
\noindent
The proof of Proposition (\ref{thm:msd}) depends on the next lemma.
\begin{lem}\label{lem:LEMMA1}
Given $\alpha\in (0,2)$,  let  $Q_{n}:=\sum_{j=0}^{n-1} A_j j^{2}$ be
the contribution to the mean square displacement of the sub-travelling areas. Then:
\begin{equation} \label{eq:tesiLEMMA1}
\lim_{n \to \infty} \frac{Q_{n}}{n^{\gamma}}=
\left\{
\begin{array}{rl}
\infty \              & \ \ \mbox{if } 0\leq\gamma<2-\alpha \\
\frac{\alpha}{2-\alpha} & \ \ \mbox{if } \gamma = 2 - \alpha \\
0                     & \ \ \mbox{if } 2 - \alpha < \gamma \leq 2
\end{array}
\right.
\end{equation} 
\end{lem}
\proof
The series $Q_{n}$ assumes a different form depending on wheter $n$ is even or odd.
If it is even and larger than $2$,
\begin{equation}
Q_{n}=\sum_{j=0,j\in P_n}^{n-1}A_j j^{2} =4\sum_{j=1}^{\frac{n}{2}-1}\left[\frac{1}{\left(2j+2^{1/\alpha}-1\right)^{\alpha}}-\frac{1}{\left(2j+2^{1/\alpha}+1\right)^{\alpha}}\right]\;j^{2}
\end{equation}
This sum has a telescopic structure that allows us to rewrite it as follows: 
\begin{equation}\label{eq:REDSUM}
Q_{n}=4~\sum_{j=1}^{\frac{n}{2}-1}\frac{2j-1}{\left(2j-1+2^{1/\alpha}\right)^{\alpha}}-
\frac{\left(n-2\right)^{2}}{\left(n-1+2^{1/\alpha}\right)^{\alpha}}.
\end{equation}
Let $R_{n}$ be the first term of $Q_n$.  Introducing $f(j):=\frac{2j-1}{\left(2j-1+2^{1/\alpha}\right)^{\alpha}}$ we can write:
\begin{equation}
R_{n}=4~\sum_{j=1}^{\frac{n}{2}-1}\frac{2j-1}{\left(2j-1+2^{1/\alpha}\right)^{\alpha}}
= 4~\sum_{j=1}^{\frac{n}{2}-1} f(j).
\end{equation}
The derivative
\begin{equation}
f'(j)=\frac{2\left[2(1-\alpha)j+2^{1/\alpha}+\alpha-1\right]}{(2j+2^{1/\alpha}-1)^{\alpha+1}}
\end{equation}
shows that $f$ is increasing for $0<\alpha\leq1$, while for $1<\alpha<2$, 
$f$ grows for $j<j(\alpha)$ 
and decreases for $j>j(\alpha)$, with $j(\alpha)=(1-\alpha-2^{1/\alpha})/2(1-\alpha)$.
For $0<\alpha \le 1$, $f$ is strictly increasing, hence:
\begin{equation}\label{eq:DIS1}
\int_{0}^{\frac{n}{2}-1}f(x)dx\leq\sum_{j=1}^{\frac{n}{2}-1}f(j)\leq\int_{1}^{\frac{n}{2}}f(x)dx.
\end{equation}
We have to distinguish two cases: $\alpha<1$ and $\alpha=1$. In the first case we have:
\begin{eqnarray}
\int_{0}^{\frac{n}{2}-1}f(x)dx &=& \frac{1}{2}\left[\frac{(n-3+2^{1/\alpha})^{2-\alpha}}{2-\alpha}-2^{1/\alpha}\cdot\frac{(n-3+2^{1/\alpha})^{1-\alpha}}{1-\alpha} \right.+ \nonumber \\
&+& \left. \left(2^{1/\alpha}-1\right)^{1-\alpha}\cdot\frac{2^{1/\alpha}-\alpha+1}{(2-\alpha
)(1-\alpha)}\right]
\end{eqnarray}
and
\begin{eqnarray}
\int_{1}^{\frac{n}{2}}f(x)dx &=& \frac{1}{2}\left[\frac{(n-1+2^{1/\alpha})^{2-\alpha}}{2-\alpha}-2^{1/\alpha}\cdot\frac{(n-1+2^{1/\alpha})^{1-\alpha}}{1-\alpha} \right.+ \nonumber \\
&+& \left.(2^{1/\alpha}+1)^{1-\alpha}\cdot\frac{(2^{1/\alpha}+\alpha-1)}{(2-\alpha)(1-\alpha)}\right]
\end{eqnarray}
therefore taking the $n\rightarrow\infty$ limit we have:
\begin{equation}\label{eq:LIMINT1}
\lim_{n \to \infty} \frac{1}{n^{\gamma}}\int_{0}^{\frac{n}{2}-1}f(x)dx=
\left\{
\begin{array}{rl}
\infty                   &       \mbox{if } 0\leq\gamma<2-\alpha \\
\frac{1}{2(2-\alpha)}    &       \mbox{if } \gamma =2-\alpha \\
0                        &       \mbox{if } 2-\alpha<\gamma\leq2
\end{array}
\right.
\end{equation}
and
\begin{equation}\label{eq:LIMINT2}
\lim_{n \to \infty} \frac{1}{n^{\gamma}}\int_{1}^{\frac{n}{2}}f(x)dx=
\left\{
\begin{array}{rl}
\infty                   &       \mbox{if } 0\leq\gamma<2-\alpha \\
\frac{1}{2(2-\alpha)}    &       \mbox{if } \gamma =2-\alpha \\
0                        &       \mbox{if } 2-\alpha<\gamma\leq2
\end{array}
\right.
\end{equation}
For $\alpha=1$ the two integrals differ, but the bounding 
limits coincide, therefore, one obtains:
\begin{equation} \label{sopra}
\lim_{n \to \infty} \frac{R_{n}}{n^{\gamma}}=
\lim_{n \to \infty} \frac{4}{n^{\gamma}} \sum_{j=1}^{\frac{n}{2}-1}f(j)=
\left\{
\begin{array}{rl}
\infty                   &       \mbox{if } 0\leq\gamma<2-\alpha \\
\dfrac{2}{2-\alpha}    &       \mbox{if } \gamma =2-\alpha \\
0                        &       \mbox{if } 2-\alpha<\gamma\leq2
\end{array}
\right.
\end{equation}

\noindent
If $1<\alpha< 2$, $f$ deacreses for $j>j(\alpha)$ hence, introducing
$\bar{j}_{\alpha}=\left\lfloor j(\alpha)\right\rfloor$, where $\left\lfloor x\right\rfloor$ 
is the integer part of $x$, $R_n$ can be expressed as: 
\begin{equation}\label{eq:REDSUM2}
R_{n}=4\left(\sum_{j=1}^{\bar{j}_{\alpha}}f(j)+\sum_{j=\bar{j}_{\alpha}+1}^{\frac{n}{2}-1}f(j)\right)
\end{equation}
Dividing by $n^\gamma$ and taking the $n \to \infty$ limit, the first term vanishes for all $\gamma>0$, while
the second term can be treated as above, to obtain the same as Eq.(\ref{sopra}).
Recalling Eq.(\ref{eq:REDSUM}), this eventually implies (\ref{eq:tesiLEMMA1}). For odd $n$, one proceeds similarly. 
\endproof
\noindent
\begin{oss} \label{eq:MSDTREND}
For $\alpha \in (0,2)$, Proposition (\ref{thm:msd}) states that the mean square displacement goes like
$\langle \Delta \hat{X}^{2}_{n}\rangle \sim n^{2-\alpha}$.
Thus, the trivial map $S_{\alpha}$ enjoys all possible regimes of standard and
anomalous diffusion, as $\alpha$ varies in $(0,2)$.
\end{oss}

The next proposition concerns the transport behavior for $\alpha=2$.  
\begin{prop}
Given $\alpha=2$ and the uniform initial distribution in $\widehat{M}_0$, 
we have:
%
\begin{equation} \label{eq:GDIFFCO2}
T_{2}(\gamma)=
\left\{
\begin{array}{rl}
+\infty               & \mbox{if }\ \  \gamma=0 \\
0                     & \mbox{if } \ \  0<\gamma\leq 2               
\end{array}
\right. 
\end{equation}
More precisely, the dynamics is \emph{``logarithmically diffusive''}, i.e.\
\[
\langle\Delta \hat{X}^{2}_{n}\rangle \sim \log n~, ~~ \mbox{asymptotically in $n$.}
\]
\end{prop}
\proof
Repeat the previous reasoning with $\alpha=2$ and correspondingly different integrals.
\endproof
\noindent
We can now state the following theorem which gives the asymptotic behavior of the moments of 
the displacement $\Delta \hat{X}^{p}_{n}$, i.e.\ of the moments of $\rho^{G}_{n}$:
\begin{equation}
\langle  \Delta \hat{X}^{p}_{n} \rangle = \sum_{j=-n}^{n} A_{j}j^{p}
\end{equation}
\begin{thm}\label{thm:MOM}
For $\alpha\in\left(0,2\right]$ the moments $\langle \Delta \hat{X}^{p}_{n}\rangle$ with $p>2$ even and initial 
condition uniform in $\widehat{M}_0$ have the following asymptotic beahvior:
\[
\langle \Delta \hat{X}^{p}_{n}\rangle \sim n^{p-\alpha}~,
\]
while the odd moments ($p=1,3,...$)  vanish.
\end{thm}
\proof
We want to compute the following limit:
\begin{equation}
L(\alpha,p):=\lim_{n\to\infty}\frac 1 {n^{\gamma}} {\langle \Delta \hat{X}^{p}_{n}\rangle}=
\lim_{n\to\infty}\ \frac 1 {n^{\gamma}}\sum_{j=-n}^{n} A_j j^{p}.
\end{equation}
As observed in Remark 2, the symmetry of $\rho^{G}_{n}$ leads the sums with odd $p$ to vanish.  
For the even moments it suffices to consider the positive $j$'s:
\begin{equation}\label{eq:AV+AQP-MOM}
L(\alpha,p) = \lim_{n \to \infty} \frac{2}{n^{\gamma}}\left(\sum_{j=0}^{n-1} A_j j^{p} + A_n  n^{p}\right).
\end{equation}
In next Lemma \ref{lem:LEMMA2} it is shown that:
\begin{equation} \label{eq:LIMAQP-MOM}
\lim_{n \to \infty} \frac{1}{n^{\gamma}}\sum_{j=0}^{n-1}A_j j^{p}=
\left\{
\begin{array}{rl}
\infty \                  & \ \ \mbox{if } 0\leq\gamma<p-\alpha \\
\dfrac{\alpha}{p-\alpha}>0 & \ \ \mbox{if } \gamma = p - \alpha \\
0                         & \ \ \mbox{if } \gamma > p - \alpha 
\end{array}
\right.
\end{equation}
while for the travelling area (\ref{eq:AV}) we have:
\begin{equation}
\lim_{n\to \infty} A_n n^{p}=\lim_{n\to \infty}\frac{n^{p}}{(n+2^{1/\alpha}-1)^{\alpha}}\cdot \frac{1}{n^{\gamma}}=
\left\{
\begin{array}{rl}
\infty      &       \mbox{if } 0\leq\gamma<p-\alpha \\
1             &       \mbox{if } \gamma = p-\alpha \\
0             &       \mbox{if } \gamma > p - \alpha .
\end{array}
\right.
\end{equation}
Therefore, we conclude that
\begin{equation}
L(\alpha,p) = 
\left\{
\begin{array}{rl}
\infty                      &      \mbox{if } 0\leq\gamma<p-\alpha \\
\dfrac{p}{p-\alpha}   &      \mbox{if } \gamma = p-\alpha    \\
0                           &      \mbox{if } \gamma > p - \alpha
\end{array}
\right.
\end{equation}
so that the large $n$ behavior of the even moments is given by
$\langle \Delta \hat{X}^{p}_{n}\rangle \sim n^{p-\alpha}$.
\endproof

\begin{lem}\label{lem:LEMMA2}
For the sub-travelling areas and any $\alpha\in (0,2]$, the following holds:
\begin{equation} 
\lim_{n \to \infty} \frac 1 {n^{\gamma}} \sum_{j=0}^{n-1} A_j j^{p} =
\left\{
\begin{array}{rl}
\infty \                & \ \ \mbox{if } 0\leq\gamma<p-\alpha \\
\dfrac{\alpha}{p-\alpha} & \ \ \mbox{if } \gamma = p - \alpha \\
0                       & \ \ \mbox{if } \gamma > p - \alpha.
\end{array}
\right.
\end{equation}
\end{lem}

\proof
Let us begin with even $n$. Defining $\mathcal{P}_{n}$ as
\begin{equation}\label{eq:SERIEMOM}
\mathcal{P}_{n}:=\sum_{j=0,j\in P_n}^{n-1}A_j j^{p} =
2^{p}\sum_{j=1}^{\frac{n}{2}-1}
\left[\frac{1}{\left(2j+2^{1/\alpha}-1\right)^{\alpha}}-\frac{1}{\left(2j+2^{1/\alpha}+1\right)^{\alpha}}\right]\;j^{p}
\end{equation}
induction leads to:
\begin{equation}\label{eq:REDSUM-MOM}
\mathcal{P}_{n}=2^{p}\cdot\sum_{j=0}^{\frac{n}{2}-2}\frac{(j+1)^{p}-j^{p}}{\left(2j+1+2^{1/\alpha}\right)^{\alpha}}-
\frac{\left(n-2\right)^{p}}{\left(n-1+2^{1/\alpha}\right)^{\alpha}} = R_{n}-
\frac{\left(n-2\right)^{p}}{\left(n-1+2^{1/\alpha}\right)^{\alpha}}~,
\end{equation}
which defines $R_{n}$ in terms of addends of the form:
\begin{equation}
f(j)=\frac{(j+1)^{p}-j^{p}}{(2j+1+2^{1/\alpha})^{\alpha}}=
\sum_{k=1}^{p}\binom{p}{k}\frac{j^{p-k}}{(2j+1+2^{1/\alpha})^{\alpha}}
\end{equation}
with derivative given by:
\begin{equation}
f'(j)=
\sum_{k=1}^{p}\binom{p}{k}\frac{\left[2(p-k-\alpha)j+(p-k)(1+2^{1/\alpha})\right]}{(2j+1+2^{1/\alpha})^{\alpha+1}}j^{p-k-1}
= \sum_{k=1}^{p}f_{k}(j)~.
\end{equation}
where
\begin{equation}
f_{k}(j)=\binom{p}{k}\frac{\left[2(p-k-\alpha)j+(p-k)(1+2^{1/\alpha})\right]}{(2j+1+2^{1/\alpha})^{\alpha+1}}j^{p-k-1}
\end{equation}
For $0<\alpha\leq 1$ and all $j>0$, we have
$f_{k}(j)>0$ for $k=1,\ldots,p-1$, while $f_{p}(j)<0$. Because $| f_p(j) | < f_1(j)$,
$f'$ is positive and $f$ increases for all $j>0$.
For $1<\alpha\leq 2$ and $p=3$, one has $f(j)=(3 j^2 + 3 j +1)/(2 j + 1 + 2^{1/\alpha})^\alpha$,
which is increasing for $j>0$, while for $p \ge 4$ one obtains 
$f_{k}(j)>0$ for $k=1,\ldots,p-2$, and $f_{p-1}(j), f_{p}(j)<0$. Because
$| f_{p-1}(j) + f_p(j) | < f_1(j) + f_2(j)$, $f(j)$ is increasing for $j>0$,
even for $1<\alpha\leq 2$.

Therefore, we can bound our sum from above and below as follows:
\begin{equation}\label{eq:DIS1-MOM}
\int_{0}^{\frac{n}{2}-2}f(x)dx < \sum_{j=0}^{\frac{n}{2}-2}f(j) < \int_{1}^{\frac{n}{2}-1}f(x)dx.
\end{equation}
for all  $\alpha \in (0,2]$.
Passing to the limit as previously done, we eventually obtain: 
\begin{equation}
\lim_{n\to\infty}\frac{R_{n}}{n^{\gamma}}=
\left\{
\begin{array}{rl}
\infty                   &       \mbox{if } 0\leq\gamma<p-\alpha \\
\dfrac{p}{2^{p}(p-\alpha)}  &       \mbox{if } \gamma =p-\alpha \\
0                        &       \mbox{if } \gamma>p-\alpha
\end{array}
\right. \quad
\lim_{n \to \infty} \frac{\mathcal{P}_{n}}{n^{\gamma}}  =
\left\{
\begin{array}{rl}
\infty                  &     \mbox{if } 0\leq\gamma<p-\alpha \\
\dfrac{\alpha}{p-\alpha} &     \mbox{if } \gamma = p-\alpha    \\
0                       &     \mbox{if } \gamma>p-\alpha
\end{array}
\right.
\end{equation}
for all $\alpha\in (0,2]$. If $n$ is odd, one proceeds similarly to obtain the same result.

\endproof

\section{Examples}
In this section, we illustrate numerically the transport produced by the Slicer Map $S_{\alpha}$ for
two different values of $\alpha$. The initial condition uniform in $\widehat{M}_0$ is approximated by 
$N$ points uniformly distributed at random on $\widehat{M}_0$, so that
the mean square displacement is approximated by:
\begin{equation}
\langle \Delta\widehat{X}_{n}^{2}\rangle \approx \frac{1}{N}\sum_{i=1}^{N}\ d_{\widehat{M}}
\left(S_\alpha^n\widehat{X}_{0}(i),\widehat{X}_{0}(i)\right)^{2}
\end{equation}
where $\widehat{X}_{0}(i)$ is the $i$-th point initial position and $d_{\widehat{M}}(\widehat{X},\widehat{Y})=k$ if $\widehat{X} \in \widehat{M}_m$ and
$\widehat{Y} \in \widehat{M}_{m \pm k}$.


\begin{ex} $\alpha=1/2$: in this case $\ell_j(1/2)={1}/\big(\left|j\right|+4\big)^{1/2}$, and
the asymptotic behavior is 
$\left\langle \Delta \widehat{X}^{2}_{n}\right\rangle \sim n^{{3}/{2}}$, cf.\ Remark (\ref{eq:MSDTREND}),
which means that $S_{1/2}$ is super-diffusive with $\gamma^{t}=3/2$, and generalized diffusion coefficient 
$T_{1/2} = {8}/{3}$, as illustrated by Fig.\ref{fig:COEFF44BIS}. Moreover, for $p>2$ Theorem \ref{thm:MOM} 
implies that:
\begin{equation}
\langle \Delta \widehat{X}^{p}_{n}\rangle \sim n^{p-{1}/{2}}
\end{equation}
For the coarse grained distribution, cf.\ (\ref{eq:DENSFUNKP}) and (\ref{eq:DENSFUNKD}), even $n$ yields:
\begin{equation}
\rho^{G}_{n}\left(m\right)=
\left\{
\begin{array}{rl}
\frac{1}{2}-\frac{1}{\sqrt{5}},       & \ \ \mbox{for } m=0\\
\frac{1}{\sqrt{2k+3}}-\frac{1}{\sqrt{2k+5}}, & \ \ \mbox{for } m=2k,\ k=2,\ldots,\frac{n-2}{2}\\
\frac{1}{\sqrt{n+3}}        & \ \ \mbox{for } m=n\\
0,                & \ \ \mbox{otherwise}
\end{array}
\right.
\end{equation}
while odd $n$ yields:
\begin{equation}
\rho^{G}_n\left(m\right)=
\left\{
\begin{array}{rl}
\frac{1}{\sqrt{2k+4}}-\frac{1}{\sqrt{2k+6}}, & \ \ \mbox{for } m=2k+1,\ k=2,\ldots,\frac{n-3}{2}\\
\frac{1}{\sqrt{n+3}}       & \ \ \mbox{for }  m=n\\
0,                & \ \ \mbox{otherwise}
\end{array}
\right.
\end{equation}
Figure \ref{fig:DENS4CONF} shows the numerically computed marginal probability 
distribution function $\rho^{G}_{n}(m)$ at a fixed even $n$, for $m>0$, including the last
value $\rho^{G}_{n}(n)$, which is much larger than the values for $m$ close to $n$ 
(the negative branch of the distribution can be recovered by symmetriy).
Because asymptotically $\rho^{G}_{n}$ goes like:
\begin{equation}\label{eq:ROX}
\rho^{\alpha}_{n}(m)=
\left \{
\begin{array}{rr}
\frac{\displaystyle C_{\alpha}}{\displaystyle ( m+2^{1/\alpha})^{\alpha+1}}~, & m<n,\\
0 ~, & m>n,
\end{array}
\right.
\end{equation} 
where $C_{\alpha}$ is a normalization constant, Figure \ref{fig:DENS4CONF} compares the numerical values 
of $\rho^{G}_{n}$ with $\rho^{1/2}_{n}(m)$, with $C_{1/2}=1$. Apart from $\rho^{G}_{n}(n)$, $\rho^{1/2}_{n}$
perfectly fits the real distribution. 
\begin{figure}[h]
\centering
\includegraphics[scale=0.5]{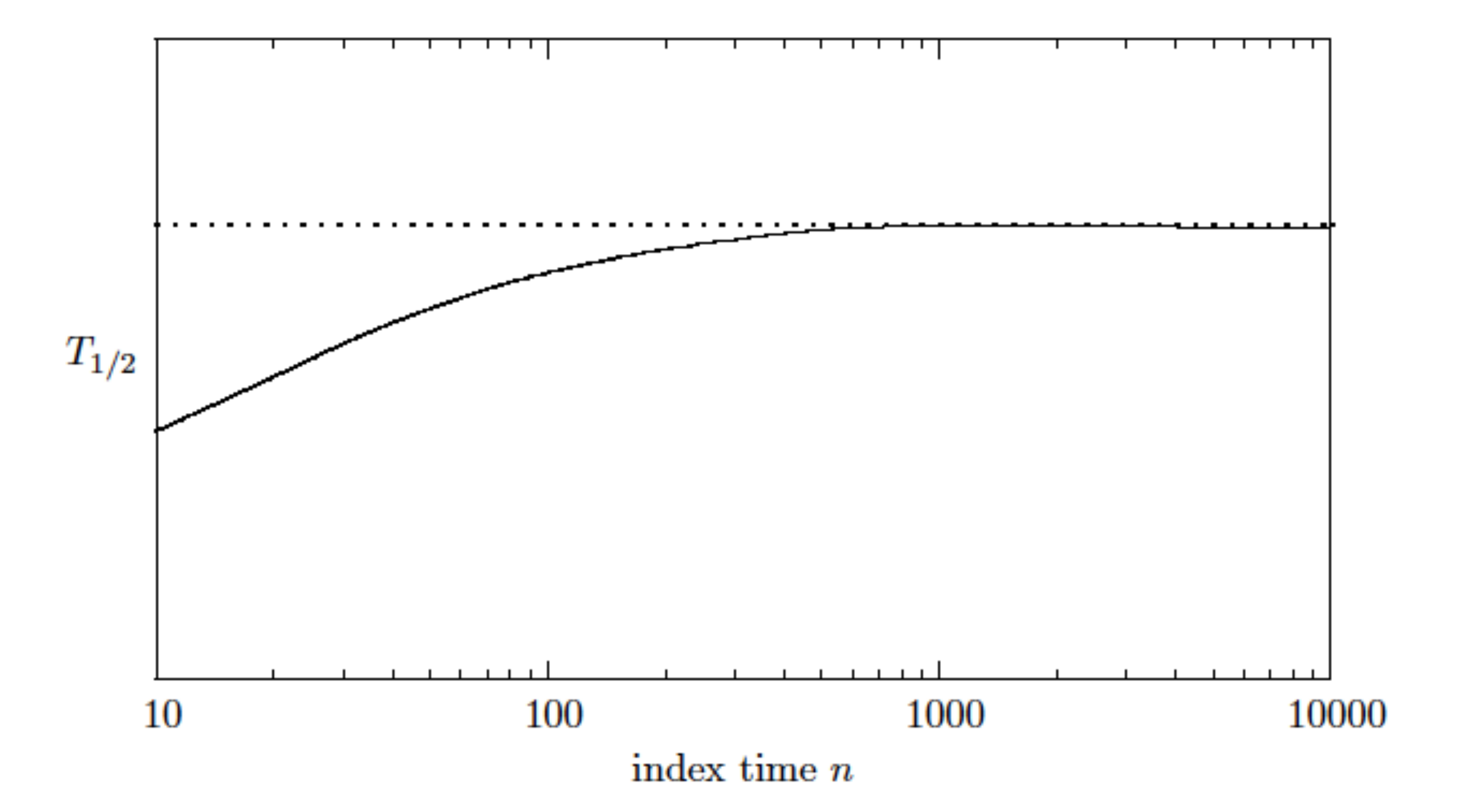}
\caption{Behavior of the numerically estimated coefficient $T_{1/2}$ (continuous line) for $n=10^{4}$ and for a choice of $N=10^{3}$ points  compared with its theoretical value  $\frac{8}{3}$ (dotted line) obtained from (\ref{eq:GDIFFCO1}).}
\label{fig:COEFF44BIS}
\end{figure}
\begin{figure}[h]
\centering
\includegraphics[scale=0.5]{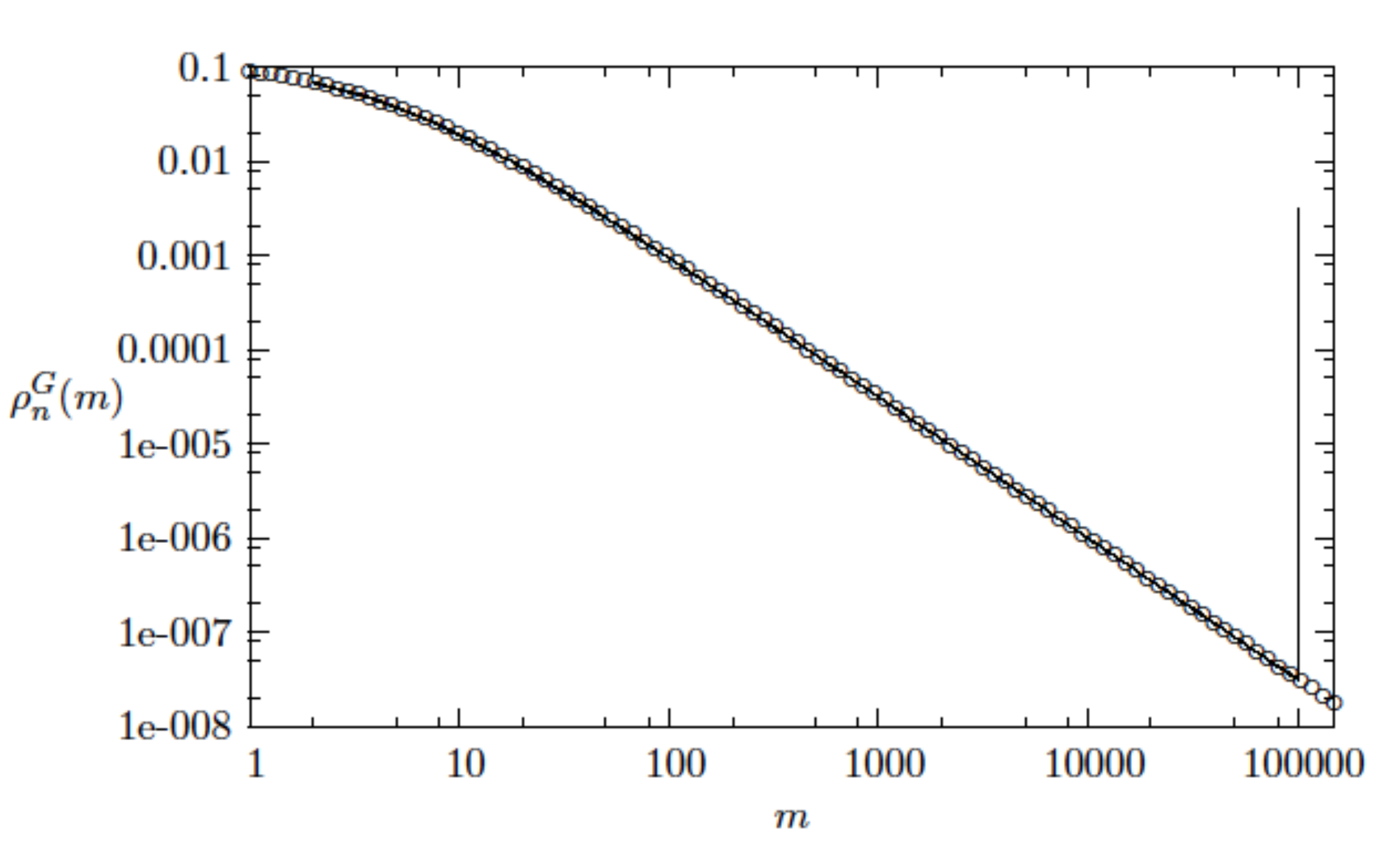}
\caption{ Log-log plot of the marginal probability distribution function $\rho_{n}^{G}(m)$ (continuous line) 
for map $S_{1/2}$ compared with $\rho^{1/2}(x)=C_{1/2}/(x+4)^{3/2}$ (circles) at fixed $n=10^{5}$. 
The peak at $m=n$ is produced by the traveling area.}
\label{fig:DENS4CONF}
\end{figure}
\end{ex}

\vskip 5pt
\begin{ex} $\alpha=1/3$: this is the case illustrated by Fig.(\ref{fig:MAPPA7FIG}), for which we have
$\ell_j(1/3)={1}/{\left(\left|j\right|+8\right)^{1/3}}$, and asymptotic behavior given by
$\langle \Delta \widehat{X}^{2}_{n}\rangle \sim n^{{5}/{3}}$. This means that
$S_{1/3}$ is super-diffusive with $\gamma^t=5/3$, and generalized diffusion coefficient $T_{1/3} ={12}/{5}$ 
(cf. Fig. \ref{fig:COEFF7}). From Theorem \ref{thm:MOM}, the moments of 
$S_{1/3}$ higher than the second have the following behavior:
\begin{equation}
\langle \Delta \widehat{X}^{p}_{n}\rangle \sim n^{p-{1}/{3}}
\end{equation}
The coarse grained distribution, see (\ref{eq:DENSFUNKP}) and (\ref{eq:DENSFUNKD}), for even $n$ reads:
\begin{equation}
\rho^{G}_{n}\left(m\right)=
\left\{
\begin{array}{rl}
\frac{1}{2}-\frac{1}{\sqrt[3]{9}},       & \ \ \mbox{for } m=0\\
\frac{1}{\sqrt[3]{2k+7}}-\frac{1}{\sqrt[3]{2k+9}}, & \ \ \mbox{for } m=2k,\ k=2,\ldots,\frac{n-2}{2}\\
\frac{1}{\sqrt[3]{n+7}}        & \ \ \mbox{for } m=n\\
0,                & \ \ \mbox{otherwise}
\end{array}
\right.
\end{equation}
while for odd $n$ we have:
\begin{equation}
\rho^{G}_{n}\left(m\right)=
\left\{
\begin{array}{rl}
\frac{1}{\sqrt[3]{2k+8}}-\frac{1}{\sqrt[3]{2k+10}}, & \ \ \mbox{for } m=2k+1,\ k=2,\ldots,\frac{n-3}{2},\\
\frac{1}{\sqrt[3]{n+7}}       & \ \ \mbox{for } m=n\\
0,                & \ \ \mbox{otherwise}
\end{array}
\right.
\end{equation}
Figure \ref{fig:DENS7CONF} compares the marginal probability distribution function 
$\rho^{G}_{n}(m)$ with $\rho^{1/3}_{n}$ given by (\ref{eq:ROX}). 
Apart from the value $\rho^{G}_{n}(n)$, the asymptotic behaviours coincide once $C_{1/3}$ is fixed.
\begin{figure}[h]
\centering
\includegraphics[scale=0.5]{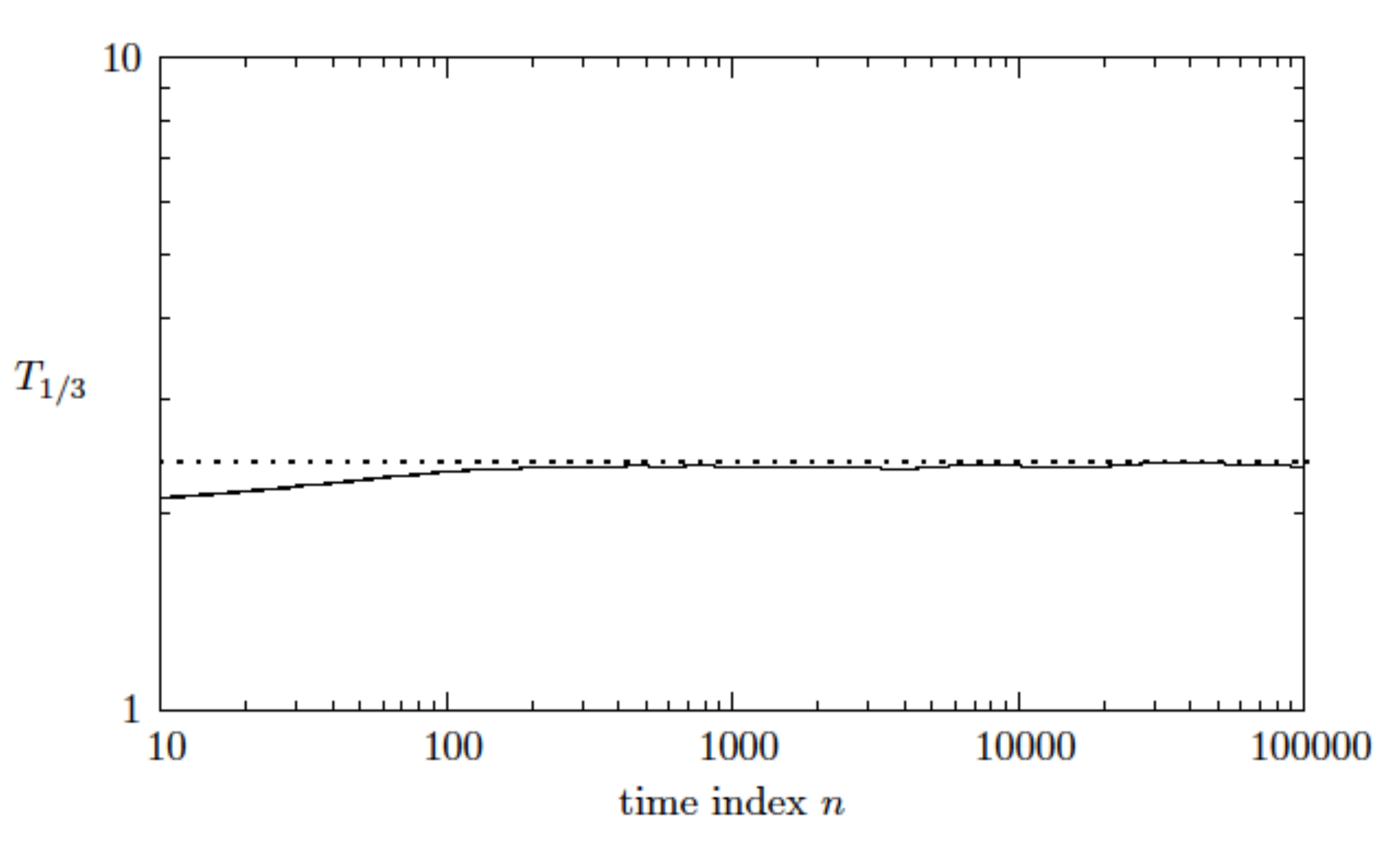}
\caption{Numerical estimate of $T_{1/3}$ (continuous line) for $n=10^{5}$ 
time steps and $N=10^{5}$ points, compared with its theoretical value ${12}/{5}$ (dotted line)
obtained from (\ref{eq:GDIFFCO1}).}
\label{fig:COEFF7}
\end{figure}
\begin{figure}[h]
\centering
\includegraphics[scale=0.5]{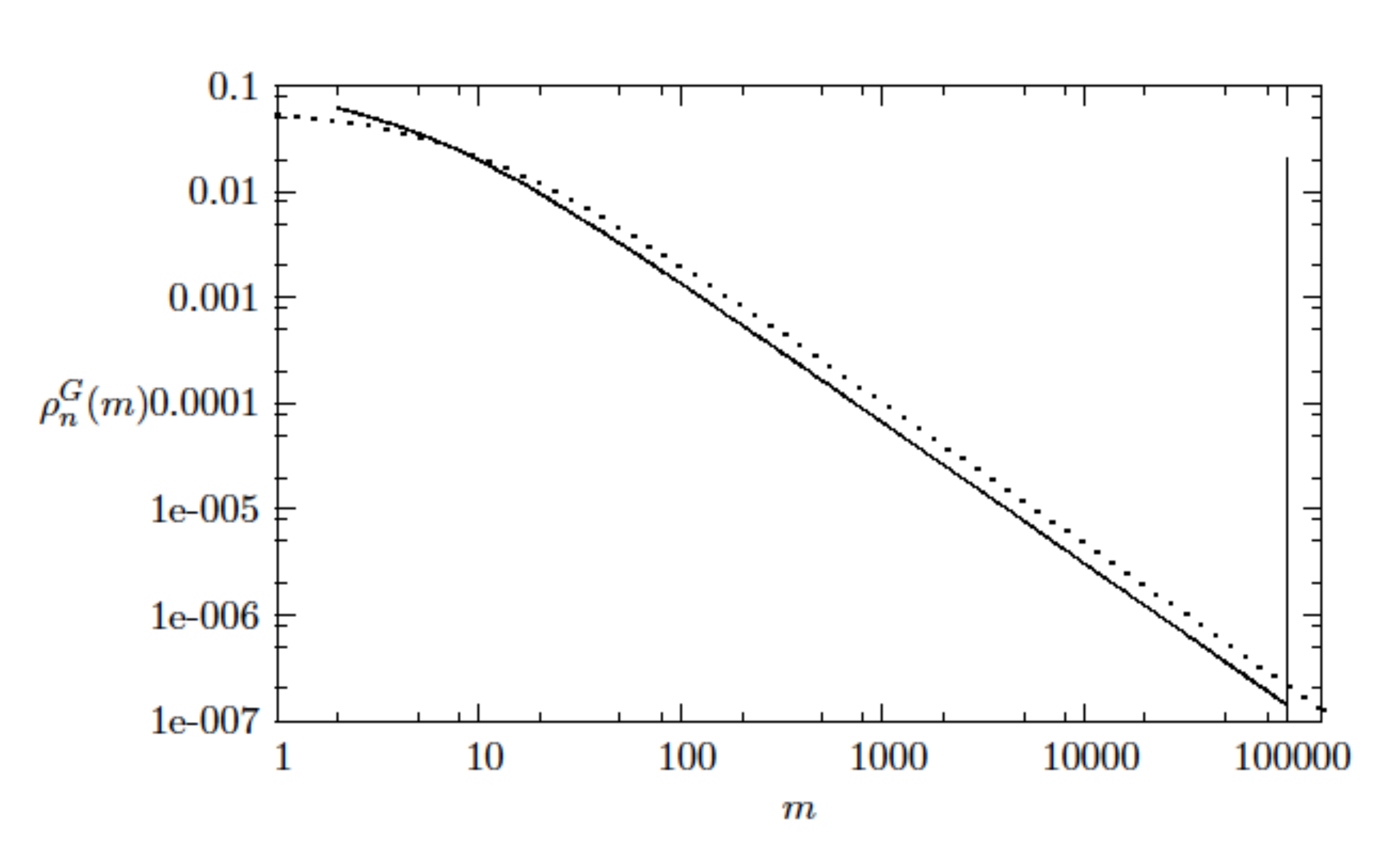}
\caption{Log-log pot of the marginal probability distribution function $\rho_{n}^{G}(m)$ 
for map $S_{1/3}$ and even $m$ (continuous line), compared with $\rho^{1/3}_{n}$ (dotted line) 
at fixed time $n=10^{5}$. A part from the values at $m=n$, where $\rho_{n}^{G}$ has a spike
due to the travelling area,
the asymptotic behaviours coincide once the normalization constant of $\rho^{1/3}_{t}$ is adjusted.}
\label{fig:DENS7CONF}
\end{figure}
\end{ex}

\section{Slicer Map and L\'evy Walks}
\subsection{L\'evy Walks in quenched disordered media }
In this section we compare the Slicer transport properties with those of the
L\'evy walks in quenched disordered media of Ref.\cite{BURIONI2010}, in which a 
one-dimensional sequence of scatterers obeys a L\'evy type distribution, with
probability density for two consecutive scatterers to be at distance $r$ given by:
\begin{equation}
\lambda(r)\equiv\beta r_{0}^{\beta}\frac{1}{r^{\beta+1}}, \ \ r\in\left[r_{0},+\infty\right)
\end{equation}
where $\beta>0$ and $r_{0}$ is a cutoff fixing the characteristic length scale of the system. Then, a walker 
moves ballistically (at constant velocity $v$) until it reaches one of the scatterers, from which it is 
either transmitted or reflected with probability $1/2$. 
The authors of \cite{BURIONI2010} derive an analytic expression for the asymptotic behavior of the mean 
square displacement $\langle r^{2}(t)\rangle$ when averaged over the scattering points. 
At first they introduce the most general scaling hypothesis for the probability distribution $P(r,t)$ 
for the walker to be in position $r$ at time $t$:
\begin{equation}
P(r,t)=l^{-1}(t)f\left(\frac{r}{l(t)}\right)+g(r,t)
\end{equation}
where $l(t)$ is the characteristic length of $P$ and the following is assumed:
\begin{equation}
\lim_{t\to +\infty}\int_{0}^{vt}\left|P(r,t)-l^{-1}(t)f\left({r}/{l(t)}\right)\right|dr=0 ~; \quad
\lim_{t\to +\infty}\int\left|g(r,t)\right|dr=0~. 
\end{equation}
The first integral stops at $vt$ because the random walker covers at most a distance $vt$
in a time $t$. For the same reason, the expression of the mean square displacement reads:
\begin{equation}
\langle r^{2}(t)\rangle = \int_{0}^{vt}l^{-1}(t)f\left({r}/{l(t)}\right)r^{2}dr + \int_{0}^{vt}g(r,t)r^{2}dr ~.
\end{equation}
Exploiting the equivalence with an electric problem \cite{DS1999}, the calculation of the number $N(t)$ of scattering 
sites visited by the walker in a time $t$ takes the form \cite{BURIONI2010}:
\begin{equation}\label{eq:NT}
N(t)\sim
\left\{
\begin{array}{rl}
t^{\frac{\beta}{1+\beta}}~,  & \  \mbox{if } 0<\beta<1\\
t^{1/2}~,                    & \  \mbox{if } \beta\ge 1
\end{array}
\right . 
\end{equation} 
Then the authors  make a ``single long jump'' hypothesis, neglecting the possibility of multiple 
consecutive jumps of length larger then a given size. For $r\gg l(t)$ this yields: 
\begin{equation}
P(r,t)\sim {N(t)}/{r^{1+\beta}}
\end{equation}
hence, the asymptotic (in $t$ and $r$) probability density is given by:
\begin{equation}
P(r,t)\sim t^{\frac{\beta}{1+\beta}}\frac{1}{r^{1+\beta}}~,~ \mbox{for } \beta<1; \quad
 P(r,t)\sim t^{\frac{1}{2}}\frac{1}{r^{1+\beta}}~,~ \mbox{for } \beta \ge 1 
\end{equation}
and the mean square displacement takes the form:
\begin{equation}\label{eq:MSDBUR}
\langle r^{2}(t)\rangle\sim
\left\{
\begin{array}{rl}
t^{\frac{2+2\beta-\beta^{2}}{1+\beta}}~, & \ \mbox{if } 0<\beta<1 \\
t^{\frac{5}{2}-\beta}~,                  & \ \mbox{if } 1\leq\beta\leq\dfrac{3}{2}\\
t ~,                                     & \ \mbox{if } \dfrac{3}{2}<\beta
\end{array}
\right.
\end{equation}
More generally, the asymptotic behavior of the moments $\langle r^{p}(t)\rangle$ for all $p>0$ is given by:
\begin{equation}\label{eq:MOMBUR}
\langle r^{p}(t)\rangle\sim
\left\{
\begin{array}{rl}
t^{\frac{p}{1+\beta}} ~,                        & \ \mbox{if } \beta<1,\ p<\beta \\
t^{\frac{p(1+\beta)-\beta^{2}}{1+\beta}} ~,     & \ \mbox{if } \beta<1,\ p>\beta\\
t^{\frac{p}{2}} ~,                              & \ \mbox{if } \beta>1,\ p<2\beta-1\\
t^{\frac{1}{2}+p-\beta} ~,                      & \ \mbox{if } \beta>1,\ p>2\beta-1  
\end{array}
\right.
\end{equation}

\subsection{Comparison}
For the asymptotic behavior for the Slicer map, Theorem \ref{thm:MOM}, we switch to a 
continuous time notation in order to compare with the continuous time process of  
\cite{BURIONI2010}, and we write $
\langle \Delta \widehat{X}^{p}(t) \rangle \sim t^{p-\alpha}$, for $0<\alpha\leq 2$.
Then, for given L\'evy walk parameter $\beta$, the second moments of the L\'evy walk and 
of the slicer map $S_\alpha$ asymptotically coincide if 
\begin{equation}\label{eq:PARAMREL}
\alpha =
\left\{
\begin{array}{rl}
{\beta^{2}}/{(1+\beta)} & \ \ \mbox{if } 0<\beta\leq1 \\
\beta - {1}/{2} &\ \ \mbox{if } 1<\beta\leq\frac{3}{2}\\
1 & \ \ \mbox{if } \beta>\frac{3}{2}
\end{array}
\right.
\end{equation}
The interesting fact for $\beta \in (0,3/2]$ is that the asymptotic forms of all even moments of 
$S_\alpha$ with $p>2$ coincide with those of \cite{BURIONI2010},\footnote{i.e.\ with lines 2 and 4 of
the right hand side of Eq.(\ref{eq:MOMBUR}), because lines 1 and 3 correspond to $p \le 2$.}
if $\alpha$ is taken from Eq.(\ref{eq:PARAMREL}).
In other words, fixing $\alpha$ and $\beta$ so that the 
second moments of the slicer dynamics and of the L\'evy walks of Ref.\cite{BURIONI2010} are asymptotically equal, 
all asymptotic moments of order $p>2$ coincide as well, if $\beta \in (0,3/2]$. Indeed, similar calcualtions to those
performed above show that restricting to the positive part of the chain, in order to compare with Ref.\cite{BURIONI2010}, 
the odd moments of the slicer map and those of the corresponding L\'evy walk are equivalent too. We have thus proved 
the following:
\begin{thm}
For any $\beta \in (0,3/2]$, the moments $\langle \Delta \widehat{X}^{p}(t) \rangle$ of order 
$p \geq 2$ of $S_\alpha$ and of the corresponding L\'evy walk asymptotically coincide if 
$\alpha$ is given by Eq.(\ref{eq:PARAMREL}).
\end{thm}
Therefore, provided $\beta$ and $\alpha$ are properly tuned, $S_\alpha$ is 
asymptotically indistiguishable from the L\'evy walks of Ref.\cite{BURIONI2010}, in the sense that 
the observables which can be expressed in terms of the moments asymptotically coincide. 
This is illustrated also by the asymptotic probability densities, since
those concerning $S_\alpha$ are of a L\'evy type, except for the spike at the extreme part 
of their tails. On the other hand, even the distributions reported in Ref.\cite{BURIONI2010}
show a peak at the largest distances.

This equivalence is guaranteed once the second moments are made asymptotically equal, but it is not trivial
 itself, because, for instance, the behaviour of the L\'evy processes changes for $\beta>3/2$, remaining always   
diffusive for increasing $\beta$. Also, it is to 
be noted that the equality of the moments concerns fixed times $t$ and not the time behaviour. Therefore, we
do not claim full equivalence of the deterministic and stochastic processes but, as usual in statistical 
mechanics, we have obtained equivalence up to a certain (rather accurate) level of observation.

\section{Concluding remarks}
In search for mathematically tractable models of anomalous diffusion, we have introduced $S_\alpha$, a map 
which reproduces all regimes of anomalous diffusion. For instance, $\alpha={1}/{3}$ yields: 
\begin{equation}
\langle \Delta \widehat{X}^{2}(t) \rangle\ \sim t^{\frac{5}{3}}~, \qquad 
\langle \Delta \widehat{X}^{p}(t) \rangle \sim t^{\frac{3p-1}{3}}
\end{equation}
where the $t^{5/3}$ behavior coincides with the numerically estimated asymptotic mean square displacement
of the periodic polygonal channel made of parallel walls which form 
angles of $90$ degrees \cite{JR2006,JRB2008}. 

The analogy between polygonal billiards and simple maps has been pursued introducing area-preserving 
non-chaotic dynamics, which have later been found to compare to those of the stochastic models 
frequently used in the study of anomalous transport. In particular, we have obtained the 
equivalence of the asymptotic moments with those of L\'evy walks. The infinitely many scales that characterize 
our slicer map through the family of slicers $L_\alpha$ seem to be indispensable to obtain the anomalous behavior.

Such a trivial deterministic area preserving and non-chaotic map as $S_\alpha$ seems to capture the essential
features of L\'evy walks. How it compares with billiard dynamics, apart from the second moment of the 
travelled distance, will be investigated in the future, since the higher moments have not been 
computed in billiard dynamics, except in several special cases.

\vskip 20pt \noindent
{\bf Acknowledgements:}
The authors are grateful to Raffaella Burioni and Rainer Klages for illuminating remarks. 
LR acknowledges funding from the European Research Council, 7th Framework 
Programme (FP7), ERC Grant Agreement no. 202680. The EC is not
responsible for any use that might be made of the data appearing herein.\\
CG acknowledges financial support from the MIUR through FIRB project 
``Stochastic processes in interacting particle
systems: duality, metastability and their applications'', grant n. RBFR10N90W and
the Fondazione Cassa di Risparmio Modena through the International Research 2010 project.

\end{document}